\documentclass[twocolumn,
superscriptaddress,
prl,
showpacs,
preprintnumbers,
amsmath,amssymb]{revtex4}

\usepackage{graphicx,epsf,epsfig,ulem}
\usepackage{epsfig}
\usepackage{bm}
\usepackage{subfigure}
\usepackage{color}
\usepackage{graphicx}

\begin{document}

\title{Orbital Stark effect and quantum confinement transition of donors in silicon}

\author{Rajib Rahman}
\affiliation{Network for Computational Nanotechnology, Purdue University, West Lafayette, IN 47907, USA}

\author{G. P. Lansbergen}
\affiliation{Kavli Institute of Nanoscience, Delft University of Technology, Delft, Lorentzweg 1, 2628 CJ, Delft, The Netherlands}

\author{Seung H. Park}
\affiliation{Network for Computational Nanotechnology, Purdue University, West Lafayette, IN 47907, USA}

\author{J. Verduijn}
\affiliation{Kavli Institute of Nanoscience, Delft University of Technology, Delft, Lorentzweg 1, 2628 CJ, Delft, The Netherlands}

\author{Gerhard Klimeck}
\affiliation{Network for Computational Nanotechnology, Purdue University, West Lafayette, IN 47907, USA} 
\affiliation{Jet Propulsion Laboratory, California Institute of Technology, Pasadena, CA 91109, USA}

\author{S. Rogge}
\affiliation{Kavli Institute of Nanoscience, Delft University of Technology, Delft, Lorentzweg 1, 2628 CJ, Delft, The Netherlands}

\author{Lloyd C. L. Hollenberg}
\affiliation{Center for Quantum Computer Technology, School of Physics, University of Melbourne, VIC 3010, Australia}

\date{\today} 

\begin{abstract} 
Adiabatic shuttling of single impurity bound electrons to gate induced surface states in semiconductors has attracted much attention in recent times, mostly in the context of solid-state quantum computer architecture. 
A recent transport spectroscopy experiment for the first time was able to probe the Stark shifted spectrum of a single donor in silicon buried close to a gate. 
Here we present the full theoretical model involving large-scale quantum mechanical simulations that was used to compute the Stark shifted donor states in order to interpret the experimental data. Use of atomistic tight-binding technique on a domain of over a million atoms helped not only to incorporate the full band structure of the host, but also to treat realistic device geometries and donor models, and to use a large enough basis set to capture any number of donor states. The method yields a quantitative description of the symmetry transition that the donor electron undergoes from a 3D Coulomb confined state to a 2D surface state as the electric field is ramped up adiabatically. In the intermediate field regime, the electron resides in a superposition between the states of the atomic donor potential and that of the quantum dot like states at the surface. In addition to determining the effect of field and donor depth on the electronic structure, the model also provides a basis to distinguish between a phosphorus and an arsenic donor based on their Stark signature. The method also captures valley-orbit splitting in both the donor well and the interface well, a quantity critical to silicon qubits. The work concludes with a detailed analysis of the effects of screening on the donor spectrum.          
\end{abstract} 


\maketitle

\section {I. Introduction}

A key feature behind the remarkable progress in solid-state electronics over the past years has been the ability to modulate the conductivity of semiconductor devices at will by using ensembles of dopants. As we approach the era of nano-scale electronics, dopants have yet another interesting role to play. Individual dopants at low temperatures provide 3D confinement to electrons and holes on length scales that are greater than individual atoms but usually less than that of quantum dots. These naturally occurring carrier traps not only provide access to a number of quantum phenomena typically associated with natural or artificial atoms, but also provide possibilities of wave-function engineering \cite{Rogge.NaturePhysics.2008, Bradbury.prl.2006} by classical control mechanisms with electric and magnetic fields. The homogeneity of the confining potential from one dopant to another of the same species is an added advantage over quantum dots, which are usually not identical in practice. On the other hand, the small length scales associated with dopants can make individual donor gate control difficult to achieve. Among other factors, developments in this area rely on a boost in the ability to scale down gate lengths to tens of nanometers.

Already, donors have been used in some elegant quantum computing (QC) proposals that draws upon the vast expertise of the semiconductor device industry. One particularly interesting proposal that renewed interest in the quantum mechanics of donors is the Kane qubit {\cite{Kane.nature.1998}}, which encodes quantum information in the nuclear spin of a phosphorus donor in silicon, and engineers the donor electron wave function by electrodes to manipulate information. Several other spin-offs of the Kane QC include encoding qubits in the electronic spin of the donor electron {\cite{Vrijen.pra.2000, DeSousa.pra.2004}} or in the spatial orbitals of a singly ionized molecule of two donors {\cite{Hollenberg.prb.2004.1}}. Recent schemes have also proposed the use of a bi-linear array of electron spin qubits {\cite{Hollenberg.prb.2006}} with semi-global field control {\cite{Hill.prb.2005}} to enhance scalability of the Kane QC and to incorporate quantum error correction and the associated circuitry. In addition to the promise of scalable system design, such architectures also benefit from the long spin coherence times in Si.  

The Kane qubit proposal has spurred a number of experimental efforts aimed at fabricating donor-based nano-structures and developing single atom {\cite{Schofield.prl.2003}} or ion {\cite{Jamieson.apl.2005}} implantation technologies. Some of the recently fabricated structures in the laboratory include a gated charge qubit device of two-donors {\cite{Andresen.Nanoletter.2007}}, a metallic wire of donors {\cite{Rueb.prb.2007}}, a single donor in a FinFET corner {\cite{Rogge.NaturePhysics.2008, Sellier.prl.2006}}, and a delta-doped layer of discrete dopants {\cite{Rueb.prb.2007_1}}. Recent experiments have been successful in measuring Stark shift of the hyperfine coupling of donors in Si {\cite{Bradbury.prl.2006}}, coherent oscillations of a P donor spin {\cite{Brandt.Nature.2006}}, orbital Stark effect of a donor coupled to a triangular well {\cite{Rogge.NaturePhysics.2008}}, and charge relaxation of a donor charge qubit {\cite{Andresen.Nanoletter.2007}}. The extensive on-going research efforts in this area are aimed at ultimately achieving the initialization, readout, and control of individual donor spins.

A single donor in Si in the proximity of a gate forms an important
system in quantum electronics, thus a great deal of effort has gone into
understanding the gate control of the electron wave function \cite{Smit.prb.2003, Larionov.nanotech.2000, Wellard.nanotech.2002, Fang.prb.2002, Kettle.prb.2003, Smit.prb.2004, Martins.prb.2004, Wellard.prb.2005, Friesen.prl.2005, Kandasamy.nanotech.2006, Debernardi.prb.2006, Hu.prb.2006, Calderon.prl.2006, Calderon.prb.2008, Rahman.prl.2007, Slachmuylders.apl.2008}. By
applying suitable gate voltages, the donor bound electron can be ionized
to the surface, where it is convenient to measure its spin and to
perform quantum control. The ionization process is adiabatic if the
donor is close to the interface but abrupt for donors buried deep into
the host \cite{Smit.prb.2003, Martins.prb.2004, Wellard.prb.2005}, and there exists a hybridized
regime between the two confinement extremes \cite{Martins.prb.2004, Rogge.NaturePhysics.2008}. 
Donors close to interfaces have also been studied recently in the context of quantum computing. In a digital version of the Kane qubit, Skinner et al {\cite{Skinner.prl.2003}} proposed a gate directed sub-interfacial transport mechanism of an ionized donor electron as a means of information transport. Calderon et al. has calculated typical adiabatic shuttling times of the electron between the donor and the interface, both from single {\cite{Calderon.prl.2006}} and two-valley {\cite{Calderon.prb.2008}}  effective mass, and concluded that the tunneling time can be sensitive to the donor depth from the interface. A single valley effective mass approach {\cite{Slachmuylders.apl.2008}} investigated the ionization process in the presence of metallic gates. In other works {\cite{Calderon.prb.2006, Calderon.prb.2007}}, it was suggested that entangling the laterally confined ionized electrons at the surface could offer more robust control over two-qubit operations, and may help to circumvent the J-oscillation problem encountered in entangling donor-bound electron spins {\cite{Koiller.prl.2002}}. 

In a recent experiment {\cite{Rogge.NaturePhysics.2008}}, the electric field dependent electronic structure of a donor near an interface was probed for the first time, thus demonstrating the soundness of the theoretical proposals. The experiment involved resonant tunneling through single donor states, and made use of single donors embedded in the corners of commercial FinFETs. To understand the transport data, we employed a tight-binding based large-scale device simulation involving over a million atoms, and obtained an accurate quantitative description of the donor spectrum. As a result, not only were we able to infer the depths of the donors and the electric fields they were subjected to, but also we could deduce the species of the donors from their Stark signature {\cite{Rogge.IEDM.2008}}. 

In this paper, we elaborate on the theoretical analysis of the gated surface-proximal donor system, and also offer a more comprehensive view of the quantum confinement transition observed in the FinFET measurements. In earlier works on this system, trial wave functions were employed in
a limited basis using either hydrogenic states, or restricted valley
effective mass theory. While these works are important milestones in our understanding of the system, the intuitive effective mass or hydrogenic approaches generally do not provide the precision required to test and interpret experimental data. Such EMT calculations only provide an incomplete description of the electronic structure, and are not able to capture many excited states, some of which could be probed in the experiments. In going beyond effective mass theory the Band Minima Basis method introduced in Ref \cite{Wellard.prb.2005} is able to describe excited donor levels in a large basis of conduction band states, but is not optimized for devices with linear dimensions beyond 10 nm. The tight-binding method involves a full-band-structure, and due to its large atomistic basis set, can capture most parts of the donor spectrum. A more complete description can provide correct trends of energy states and correct symmetry transitions of the wave functions particularly near the ionization regime.  



The importance of the excited states in the basis has been evident from previous EMT works on the Stark shift of the energy spectrum of donors buried deep in bulk silicon. Inclusion of the p-states in the calculation {\cite{Debernardi.prb.2006}} was shown to improve
the evolution trend of the ground state obtained from a 1s-manifold
model {\cite{Friesen.prl.2005}}. 

While most other works have been done on P donors, we have modeled P and As impurities in detail to help positively identify the impurities found in the experiment as As. We also investigate the effect of various types of screening on the donor spectrum. Such screening effects in real devices can range from purely metallic to half-metallic or insulator type, and need to be a part of any realistic donor-interface model.   

 
\begin{figure}[htbp]
 \centering
 \includegraphics[width=3.2in,height=2in]{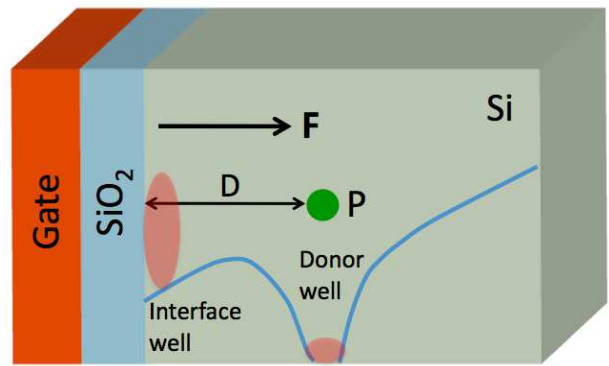}
 \caption{A schematic of a single-donor device. An electric field perpendicular to the oxide interface generates a potential well at the surafce, which can can couple to the Coulombic potential well produced by a donor. The electronic structure of the whole system is sensitive to the donor depth $D$ and the applied field $F$.}
\label{fi:1}
\end{figure}
 
A schematic of the device under investigation is shown in Fig \ref{fi:1}.  A Group V donor is located a distance D from the oxide barrier in a lattice of Si atoms. The donor generates a Coulomb potential well that traps an electron at low electric fields and at low temperatures. A uni-directional electric field is applied perpendicular to the oxide surface, and generates a triangular well at the interface. At low electric fields, the donor well is much lower in energy than the triangular well, and the lowest states of the system are localized in the donor well with symmetries permitted by a 3D Coulomb well, host band-structure and interface effects. At high electric fields, the interface well is lower in energy, and states are localized at the surface forming a 2D system. The transition from the 3D Coulomb confinement to the 2D surface states occur at intermediate field values at which the two wells are almost aligned in energy. 
  
This paper is organized as follows. In Section II, we elaborate on the details of the method. In Section III, we discuss the Stark spectrum of a donor in detail dividing the spectrum into three field regimes. Explanations of interface effects, donor species and depths, valley-splitting in interfacial states, and effects of image charges are also provided. Section IV concludes this work.

\section{II. Method}

The tight-binding method employed in this work utilizes the 20 band $sp^3d^5s*$ spin model with nearest neighbor interactions. This model is based on representing wave functions of solid-state systems with linear combination of atomic orbitals (LCAO) after the semi-empirical treatment proposed by Slater and Koster {\cite{Slater.physrev.1954}}. The model parameters were optimized by a genetic algorithm procedure {\cite{Klimeck.cmes.2002}} with analytically derived constraints {\cite{Boykin.prb.2004}} to fit critical features of the Si band-structure. This is a widely applied technique in semi-empirical tight-binding theory to model a host of semiconductor materials. 

The donor was modeled by a Coulomb potential screened by the dielectric constant of Si. The donor potential was forced to assume a cut-off potential $\textrm{U}_0$ at the donor site, the magnitude of which was adjusted to obtain the ground state binding energy of the donor. It was shown in an earlier work {\cite{Shaikh.encyclopedia.2008}} that the magnitude of $\textrm{U}_0$ approximates the strength of the valley-orbit interaction responsible for lifting the six-fold degeneracy of the 1s manifold of the donor in bulk. 
The full Hamiltonian of the host and the donor subjected to a constant electric field and closed boundary conditions was diagonalized by parallel Lanczos algorithm to extract the relevant part of the donor spectrum {\cite{Naumov.jce.2004}} with the NEMO-3D (Nano-Electronic Modeling Tool) simulation engine {\cite{Klimeck.cmes.2002, Klimeck.ted.2007}}. Since dangling bonds at the surfaces can introduce spurious eigen values at the bandgap, a model of surface passivation {\cite{Lee.prb.2004}} was employed to eliminate the states in the band gap and to improve the reliability of the eigen solver. Each of the simulations in this work typically used a 3D zincblende atomistic lattice of about 1.4 million Si atoms, and took 6 hours on 40 processors to capture 14 energy states {\cite{nanohub.note}}.   

The same technique was previously used to compute Stark shift of the donor hyperfine interaction {\cite{Rahman.prl.2007}} in good agreement with ESR experiments {\cite{Bradbury.prl.2006}}. The same model has also been applied to compute valley-splitting in quantum wells in the presence of lattice miscuts and alloy-disorder {\cite{Neerav.apl.2007}}, and to model quantum dots for optical communication wavelengths {\cite{Usman.nano.2008}}. 


\section{III. Results and Discussions}

It is well known that a Group V donor in bulk Si has an orbital singlet ground state of $\textrm{A}_1$ symmetry, an orbital triplet manifold of first excited states of $\textrm{T}_2$ symmetry, and an orbital doublet manifold of second excited states of $\textrm{E}_1$ symmetry {\cite{Kohn.physrev.1955}}. The six lowest states of the donor are of 1s type, and arise from the six-fold degenerate conduction band minima of Si. For a P donor in Si, the above three manifolds are at -45.6 mev, -33.9 meV and -32.6 meV respectively below the conduction band {\cite{Ramdas.progphys.1981}}. In addition, there are higher manifolds of notably 2p and higher states bound at approximately -11 meV. In comparison, the only notable difference in this spectrum for an As donor is the ground state energy of -54 meV instead of -45.6 meV. 

The splitting of the six 1s states of a donor into the three components described above is due to the valley-orbit (VO) interaction {\cite{Pantelides.prb.1974}}, which is the result of coupling between the conduction band valleys produced by the rapidly varying donor potential in the vicinity of the nucleus. The VO interaction varies from one donor species to another due to the species dependent microscopic variation of the donor potential in the central cell. These central cell effects are caused by a number of factors such as distance dependent dielectric screening and local strain in the bonds between the donor and the host atoms {\cite{Pantelides.prb.1974, Wellard.prb.2005}}.

\begin{figure}[htbp]
 \centering
 \includegraphics[width=3.2in,height=6in]{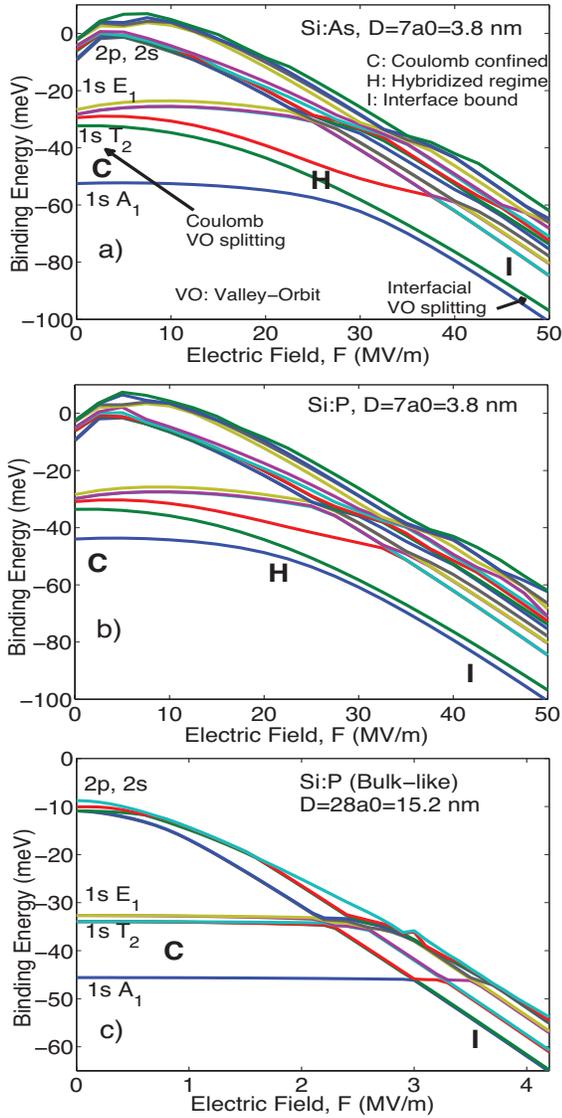}
 \caption{The electronic structure of a donor near an interface as a function of electric field. a) and b) depict the spectrum of an As and P donor respectively at a depth of 3.8 nm (7 lattice constants), while c) is for a P donor at 15 nm depth (bulk-like case). The letters C, H, and I mark the three confinement regimes: Coulomb confined (C), Hybridized been donor and interface states (H), and 2D interface confined (I). }
\label{fi:2}
\end{figure}

\begin{figure}[htbp]
 \centering
 \includegraphics[width=3.2in,height=2.8in]{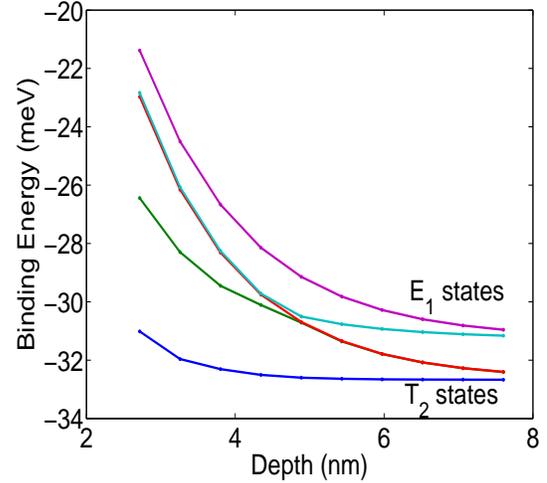}
 \caption{The orbital triplet ($\textrm{T}_2$) and the orbital doublet ($\textrm{E}_1$) manifolds as a function of donor depth. While all the states are pushed up by confinement, components of $\textrm{T}_2$ and and $\textrm{E}_1$ are seen to anti-cross each other at low depths.}
\label{fi:3}
\end{figure}

In Fig \ref{fi:2}, we show the Stark shifted spectra of donors in Si. The top panel (a) shows the spectrum for an As impurity at a depth of 3.8 nm (7 lattice units) from the interface. The middle panel (b) shows the spectrum of a P donor at the same depth, while the bottom panel (c) is for a P donor 15 nm from the interface, mimicking a bulk donor as surface effects do not influence the donor states at zero field. 
The field range is chosen such that we capture the entire transition of the donor electron from the impurity well to the interface well. The following analysis is broken down into three field ($F$) regimes. All the energies in this work are shown relative to the bulk conduction band minima along with the potential drop between the donor and the interface subtracted from it.

\begin{figure}[htbp]
 \centering
 \includegraphics[width=3.4in,height=6in]{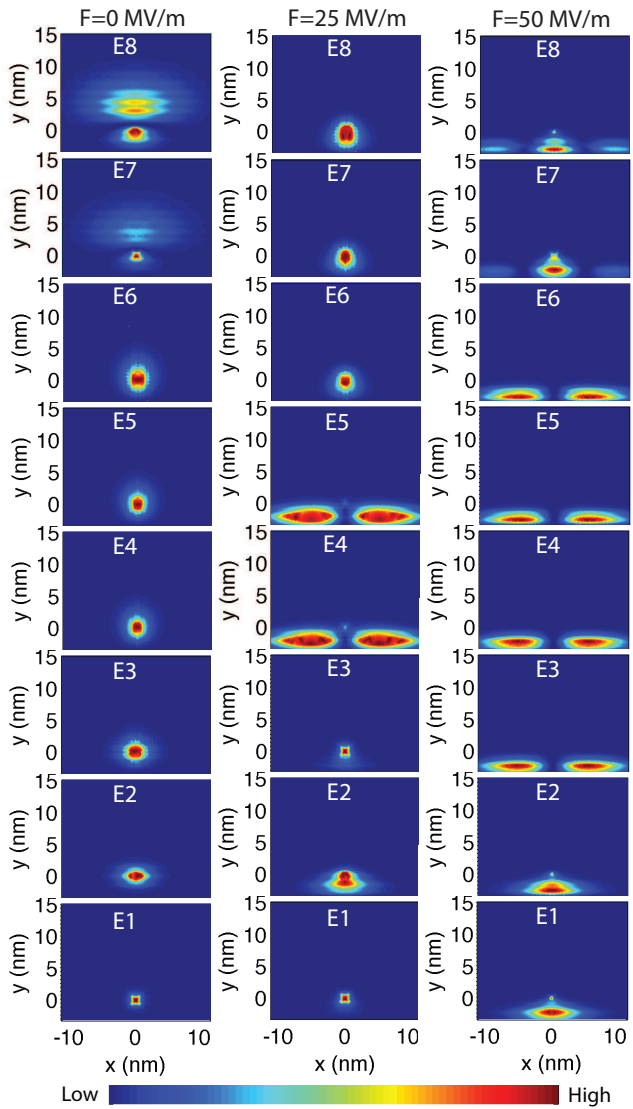}
 \caption{The lowest 8 single electron probability densities ($|\Psi|^2$)i of the Si:As system in the a) Coulomb confined regime (left column), b) intermediate field regime (middle column), and c) interfacial confinement regime (right column). The As donor is at 3.8 nm from the interface, and its energy spectrum is shown in Fig \ref{fi:2}a. The field and depth are both in the y-direction. The plots show a 2D through the $z=0$ plane passing through the donor center.}
\label{fi:4}
\end{figure}

\subsection{Low field regime}

At $F=0$, the states are all confined to the donor well. While the bulk impurity case of the bottom panel shows the singlet, triplet and doublet manifolds at the respective energies described above, an interface breaks this symmetry for a donor located close to the Si boundary. For both an As and a P donor about 3.8 nm from the interface, the degeneracy of the triplet (doublet) states is lifted. A closer look at the zero-field states as a function of donor depth, as shown in Fig \ref{fi:3}, reveals the effect of a planar interface on these Coulomb confined states. As the donor depth decreases, all the states are pushed up in energy due to confinement, similar to what is observed in a quantum well as the width of the well decreases. The triplet state is split into components of two and one, while both the doublet states are split. The two-fold degenerate component of the triplet anti-crosses one of the doublet states at about a donor depth of 5 nm (4 Bohr radii). The states are restored to their bulk symmetries at larger depths of about 7 nm (6 Bohr radii).

At low electric fields, the ground state is unaffected, while the higher states evolve downwards in energy \cite{Calderon.prl.2006, Debernardi.prb.2006}. This downward movement is more pronounced for the higher manifold of p-states. For small donor depths, the s-type excited states and the p-states exhibit a slight upward evolution before following their general trend of downward evolution in the energy scale. This could be due to the interface-induced truncation of the wave functions, giving rise to small first-order shifts in energy. A similar effect was also observed in our earlier work on the Stark shift of the contact hyperfine coupling {\cite{Rahman.prl.2007}}. We found an additional linear Stark shift contribution to the hyperfine coupling in donors located near interfaces, and could relate it to the non-zero dipole moments arising from lack of even symmetry in wave functions truncated by interfaces.

The 8 lowest wave functions are shown in Fig \ref{fi:4}. The left column shows the wave functions at $F=0$ for the As donor at 3.8 nm depth. The first six states are seen to have 1s-type symmetries although they are truncated by the interface. The 7th and 8th states are of p-type, and was also shown in Ref {\cite{Slachmuylders.apl.2008}}, although a few of the 1s states arising due to the multi-valley structure of Si were not captured in that work.  

\subsection{Intermediate field regime}

As the electric field increases, a triangular well is formed at the interface, and the higher states of the system have interfacial confinement. At intermediate field values, the interface well and the donor well is somewhat aligned in energy. At this point, the higher lying p-states and the interface states mingle with the 1s manifold (Fig \ref{fi:2}) pushing the whole manifold downwards in energy. In this regime, strong hybridization is observed between the donor states and the interface states, as the donor bound electron begins its ionization to the interface. The second excited state ($\textrm{E}_3$), which was moving downwards in tandem with the first excited state ($\textrm{E}_2$), begins to anti-cross the ground state ($\textrm{E}_1$), while $\textrm{E}_2$ continues to evolve downwards. At this point, the ground state begins to evolve downwards while $\textrm{E}_3$ moves up and mixes with the higher states. This regime marks a symmetry transition from the 3D Coulomb confined states to 2D interface states. This also serves as a signature of an atomic Coulomb well linked to a gate-generated quantum dot like structure. 

The middle column of Fig 4 shows some of the wave functions in this hybridization regime. The electron resides in a superposition of the donor state and the interfacial state, as shown in the probability densities of $\textrm{E}_1$, $\textrm{E}_2$ and $\textrm{E}_3$. States $\textrm{E}_4$ and $\textrm{E}_5$ are actually excited interface states, which penetreated the 1s manifold of the donor. $\textrm{E}_6$, $\textrm{E}_7$ and $\textrm{E}_8$ are still confined at the impurity.

\subsection{High field regime}

Increasing the electric field further pushes the interface well below the impurity well. As a result, the states are mostly localized in the interface well and has 2D symmetries. It is to be noted that the long-range Coulomb potential still binds the electron laterally at the interface and prevents it from forming a 2DEG over an extended lattice. This gives rise to the possibility of preserving identities of qubits, as pointed out in Ref {\cite{Calderon.prb.2006, Calderon.prb.2007}}, as well as producing interfacial qubits with a lesser number of gates. 

Since the uni-directional electric field lowers two of the four valleys of Si, we expect a manifold of two lowest states arising from the contribution of the lowered valleys. These two states are expected to be somewhat isolated from the higher manifold of states. In Fig \ref{fi:2}a, we observe the two closely spaced states occuring above $F=30$ MV/m. A gap of about $40$ meV with the higher manifold is also observed. States comprising of one, two and three lobes are observed in Fig 4 (right column) in this regime. The higher states $\textrm{E}_7$ and $\textrm{E}_8$ are still somewhat hyrbidized with donor states, whereas the lowest states do not have much electron density near the impurity site.

\begin{figure}[htbp]
 \centering
 \includegraphics[width=3.4in,height=1.8in]{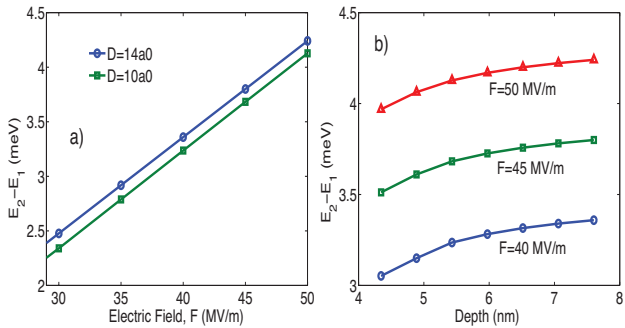}
 \caption{a) Splitting of the lowest two interfacial states as a function of a) field, and b) donor depth.}
\label{fi:5}
\end{figure}

The splitting between the two lowest states in this field regime is due to valley splitting resulting from confinement. The strong confinement potential of the hard wall interface on one end and the electric field on the other cause coupling between the two lowered valleys, and result in a splitting between the states to which these valleys contribute. This phenomena has been intensively studied in Si quantum wells and dots {\cite{Boykin.apl.2004, Neerav.apl.2007}}, where valley splitting can be engineered to separate out the spin states used for encoding qubits. 

In Fig 5a, we plot this interfacial valley-splitting as a function of electric field for two different donor depths. In the field regime shown, valley-splitting increases linearly with the field as the triangular confinement provided by the electric field becomes stronger. Fig 5b shows valley-splitting as a function of donor depth at three field values. At a constant electric field, the splitting seems to increase non-linearly with donor depths, and flattens out at higher depths. 
This is a consequence of the fact that a higher field is needed to ionize the electron bound to donors closer to the interface. While the confinement provided by the interfacial hard wall was held fixed for the data in Fig 5, we will show later that the magnitude of the valley splitting is affected by image charges that modify the interfacial confinement potential. However, the general trends of the graphs in Fig 5 with field and depth remain unchanged irrespective of the screening effects.

The presence of this interfacial valley splitting is critical for proposals in which 2D confined electrons at the surface are to be used as qubits. By increasing the field, the two states can be sufficiently isolated and quantum information can be encoded in the two-fold degenerate spin states. However, such architectures need to account for the screening dependence of the valley splitting and the spin-orbit contributions of the Rashba effect for practical operation.   

\subsection{Donor species and depths}

Comparison of the Stark shifted spectrum of an As donor with a P donor at the same depths of 3.8 nm, as shown in Fig \ref{fi:2}a and \ref{fi:2}b, reveals the basic trends of the eigenstates to be similar. The only notable difference arises in the spacing between the ground state and the excited manifold since As has a higher binding energy than P. As a consequence, the P donor states reach the hybridization regime at a lower field. A transport spectroscopy experiment, which can probe the energy spacings of a few of the excited states relative to the ground state, can determine the species of the donor with the aid of a statistical fitting procedure presented in Ref {\cite{Rogge.IEDM.2008}}. This technique, however, relies on a measurable difference between the binding energies of the group V donors, and is not likely to be successful for donor pairs like P and Sb whose binding energies only differ by less than 2 meV.    

The onset of ionization occurs when the interface well states are at similar energies to the donor states - a regime we denote as a hybridized regime since the eigenstates are in a superposition of the donor and the interface states. For larger donor depths, this hybridization occurs at lower fields as it takes a smaller field to cause the same drop in potential between the donor and the interface. In Ref {\cite{Calderon.prl.2006}}, it was shown that the critical field at which the donor and the well ground states anti-cross decreases with depth. For donors at small depths, the electron resides in a superposition state over a range of field values as its ionization is not abrupt like a bulk impurity. In our earlier work  {\cite{Rogge.NaturePhysics.2008}}, we were able to identify a hybridized regime in a field-depth curve, and map the experimental data points on this curve. A data sample to the left of this curve signified a Coulomb confined regime, whereas a data sample to the right signified an interfacial confinement regime. As the donor depth increases, the width of this hybridized regime gets narrower as the donor-interface coupling diminishes. Comparison of a P donor at 3.8 and 15.2 nm (Fig \ref{fi:2}b and \ref{fi:2}c) shows that not only does the ionization field decrease as depth increases \cite{Friesen.prl.2005, Debernardi.prb.2006, Calderon.prl.2006}, but also the field regime for hybridization becomes narrower.  
 
\subsection{Electron localization}

\begin{figure}[htbp]
 \centering
 \includegraphics[width=3.2in,height=6in]{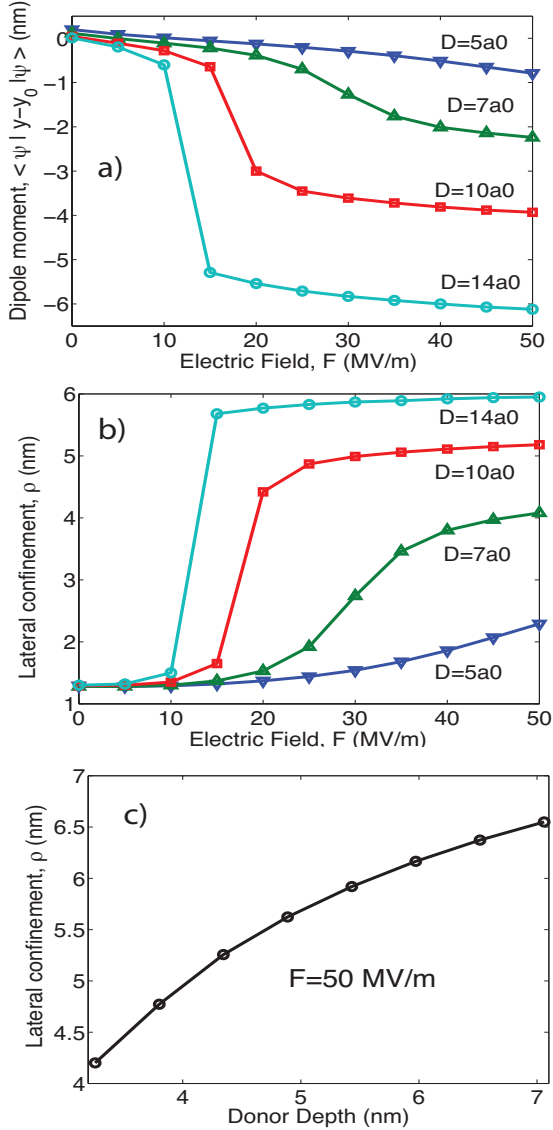}
 \caption{a) The ground state dipole moment in the direction of the field showing average electron localization. The electron shuttling is smooth for donors near the surface, but abrupt for donors buried deep. b) The lateral confinement of the electron as a function of field for different donor depths. c) The variation of lateral electron confinement at the interface as a function of donor depth at $F=50$ MV/m. All the data are for an As donor.}
\label{fi:6}
\end{figure}

Fig 6 gives a quantitative description of the electron localization at different fields and donor depths. Fig 6a shows the dipole moment in the direction of the field for different donor depths as a function of the field. At $F=0$, the electron is localized at the impurity, and the dipole moment is 0. As the field is increased, the electron probablity distribution shifts towards the interface either gradually for small donor depths or abruptly for larger donor depths. Once ionized, it exhibits a weaker dependence on the field. 

To provide some insight into how strongly the electron is laterally bound at the interface, we can make use of the expectation value of the operator $\rho=\sqrt{(x-x_0)^2+(z-z_0)^2}$ as the field is in y-direction with $(x_0, y_0, z_0)$ being the coordinates of the impurity. Fig 6b and 6c show this lateral confinement of the donor electron as a function of field and depth respectively. At $F=0$ in fig 6b, the lateral confinement is between 1 and 2 nm, which is of the order of the Bohr radii of the donor. As the field increases, the lateral confinement deteriorates as the electron moves away from the impurity core. Fig 6c shows the lateral confinement at the interface as a function of donor depth at a high field value of $F=50$ MV/m. As expected, the lateral confinement is strongest for donors close to the interface. 

This shows that experiments which are aiming to build interfacial qubits may benefit from a delta doped layer of impurities at depths chosen to suit their confinement criteria based on gate densities and qubit separations for optimal exchange interactions.   

\subsection{Effects of screening}

In realistic devices, presence of charges near a boundary between a semiconductor and another material can induce image charges. These image charges occur because of a redistribution of the charges in the vicinity of the boundary, and they affect the elctro-statics of the system by modifying the net potential the source charges experience. The image charges and their screening effects strongly depend on the materials at the other side of the boundary, most notably through their dielectric functions. 

In Ref {\cite{Macmillen.prb.1978}}, MacMillen used a variational technique to derive an approximate model of screening for a donor near an interface. Assuming that the donor is located at the coordinates $(x_0, y_0, z_0)$ and the interface is closest in the y-direction, the additional screening potential due to the image charges in his model is of the form,

\begin{equation} \label{eq:image} 
H_S=\frac{CQ}{\sqrt{(x-x_0)^2+(y+y_0)^2+(z-z_0)^2}}-\frac{CQ}{4y}
\end{equation}

\noindent
where the first term represents the interaction of the electron with the image of the positively charged nuclear core, while the second term is the interaction of the electron with its own image. In effect, the first term is that of a point charge Q located a distance D on the other side of the interface and interacting with the donor electron. The 2nd term due to the electron image term is a 1D confined potential commonly used to describe electronic image screening effects in 2DEGs. $C$ is the electrostatic constant given by $e^2/(4\pi\epsilon_{si})$. In comparison, the unscreened Hamiltonian of the system can be expressed as,

\begin{equation} \label{eq:h1} 
H_U=H_0-\frac{C}{\sqrt{(x-x_0)^2+(y-y_0)^2+(z-z_0)^2}}+eFy
\end{equation}

\noindent
where $H_0$ is the Si crystal Hamiltonian, the 2nd term is the donor potential energy, and the 3rd term represents the y-directed electric field. The total Hamiltonian is given by, $H_T=H_U+H_S$.

Although we employ this model in this work, a more accurate model may involve a self-consistent Possion solution taking into account the probability distribution of the electron. Such a model would capture the lateral confinement of the electron image missing in this work. It was also suggested in another work {\cite{Diarra.prb.2005}} that the electron image charge term assumes a more gradual variation and does not assume such a high value at the dielectric boundary. For simplicity and ease of computation, we have ignored the two above-mentioned corrections. The screening model in (1) has been used in other works {\cite{Calderon.prl.2006, Slachmuylders.apl.2008}}, and presents a good basis for comparison.      

In equation (1), $Q$ is a ratio given by $Q=\frac{\epsilon_I-\epsilon_{Si}}{\epsilon_I+\epsilon_{Si}}$. For a metallic interface, $\epsilon_I=\infty$ and $Q$ reduces to 1. An $\textrm{SiO}_{2}$ interface has $\epsilon=3.4$, and $Q$ assumes the value -0.55. $Q$ also vanishes if the interface material is Si suggesting that there are no image charges if there is no dielectric discontinuity. It is to be noted that for a metallic interface the image charges have opposite signs as the source charges, which implies that the electron image tends to pull the electron toward the interface while the donor image tends to push the electron away from the interface. For an insulator interface like $\textrm{SiO}_2$, the image charges are of the same sign as their source charges, and has the reverse screening effect as compared to a metal. Since a small layer of oxide is sandwiched between the metal and the semiconductor in realistic devices, a more realistic screening might be something between a metallic and an insulator type screening. We also investigated the screening effects for such a case with $Q=0.5$, henceforth referred to as partial metallic screening (PM).    

\begin{figure}[htbp]
 \centering
 \includegraphics[width=3.4in,height=3.4in]{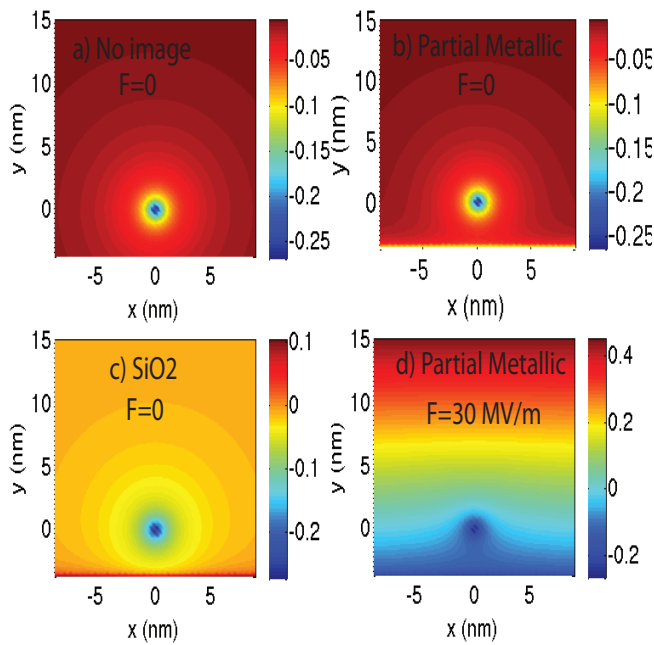}
 \caption{ Effect of screening on the donor potential.  The donor is Si:As at a depth of 3.8 nm. a) The potential at $F=0$ without any image charge effects. Total donor potential with b) partial metallic type screening ($Q=0.5$), and c) $\textrm{SiO}_{2}$ type screening ($Q=-0.55$), and d) partial metallic type screening at $F=30$ Mv/m.}
\label{fi:7}
\end{figure}

Fig 7 shows the net potential the donor electron is subjected to under different types of screening. Plot 7a ignores screening, plot 7b and 7d employ partial metallic screening, whereas plot 7c assumes insulator type screening of $\textrm{SiO}_{2}$. Plots 7a, 7b and 7c are all at zero electric fields. Comparison of 7a and 7b shows that the partial metallic type image charges cause the potential well to spread out more near the interfacial region and advocates ionization. Fig 7c shows that oxide type screening not only raises the net potential, but also provides more donor confinement and hinders ionization. Fig 7d shows the screened donor under a strong electric field. 

\begin{figure}[htbp]
 \centering
 \includegraphics[width=3.4in,height=1.8in]{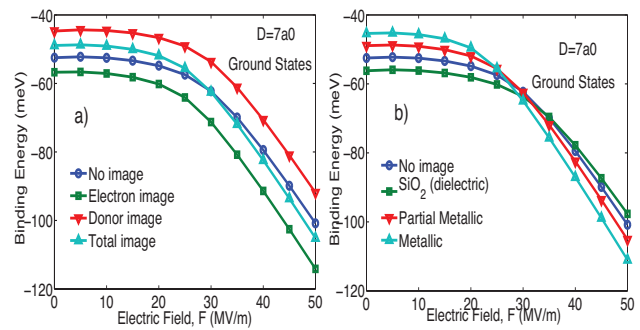}
 \caption{Effect of image charges on the ground state of Si:As  at 3.8 nm donor depth. a) Effect of the various image potential terms of eq 1 on the binding energy with partial metallic type screening. b) Variation of the binding energy with various types of screening.}
\label{fi:8}
\end{figure}

Fig 8a shows the effect of each of the image charge terms of eq (1) on the binding energy of the donor with partial metallic type screening. If the first term of eq (1) is taken into account only, the net attractive potential of the system is lowered as the donor image term is of opposite sign to the donor source potential term. Hence the binding energy of the electron decreases at all field values. On the other hand, the electron image term is attractive and increases the total attractive potential the donor electron experiences. As a result, the donor electron is more strongly bound relative to the conduction band edge. If we include both the image terms and compare the resulting binding energy with the unscreened binding energy, we notice that the binding energy decreases (less negative) in the Coulomb confined regime and increases (more negative) in the interfacial confinement regime. This suggests that the donor image term plays a dominant role in the Coulomb confined regime, while the electron image term is more dominant in the interfacial regime. There is a point at which the unscreened and screened binding energy curves cross each other, implying that the donor and the electron image effects completely cancel each other.  

In Fig 8b, we show the binding energy with various types of screening. All three screened binding energy curves cross the the unscreened binding energy curve, suggesting that the donor and electron image terms switch their dominant roles between the Coulomb and the interfacial confinement regimes. A closer look at the interfacial regime at $F=50$ MV/m shows that the $\textrm{SiO}_2$ screened curve has the lowest binding energy while the metallic screened curve has the highest. This is expected provided the electron image term plays a dominant role in the interfacial regime. In the metallic case, the electron image term is attractive and cause the electron to be strongly bound at the interface. The repulsive electron image term in case of $\textrm{SiO}_2$, causes the electron to be bound at a lower energy.

\begin{figure}[htbp]
 \centering
 \includegraphics[width=3.2in,height=6in]{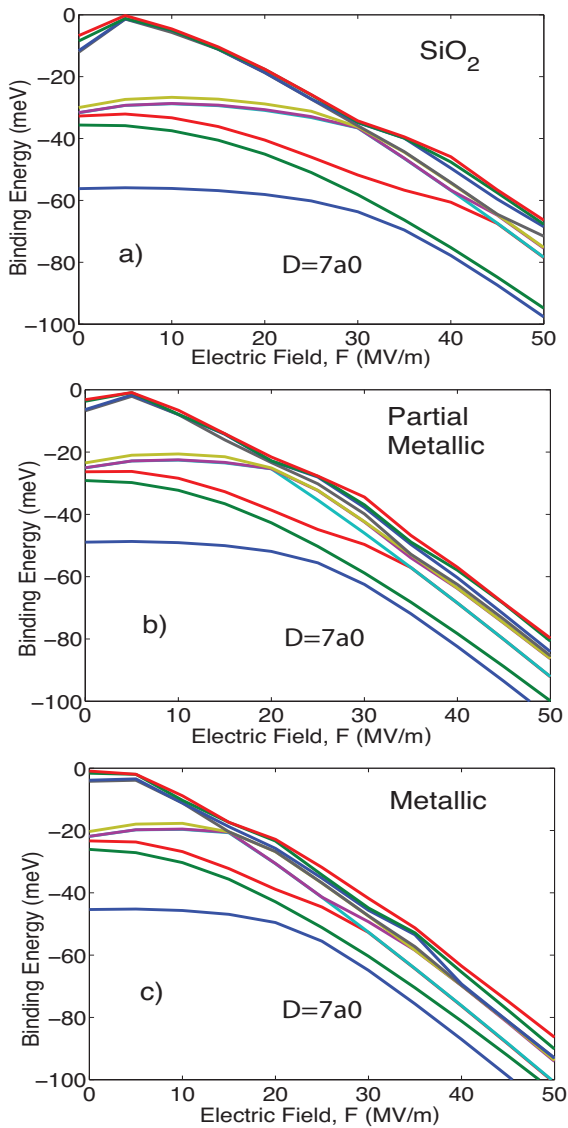}
 \caption{Partial Stark spectrum of Si:As at 3.8 nm depth with a) $\textrm{SiO}_2$ type screening ($Q=-0.55$)b) partial metallic type screening ($Q=0.5$) c) Metallic screening ($Q=1$).}
\label{fi:9}
\end{figure}

In Fig 9, we show part of the Stark shifted spectrum for the As donor at 3.8 nm depth under different types of screening. The major effect of screening is a shift of the whole spectrum in absolute energy scale. The relative differences between the lowest states in the Coulomb confined and in the hybridized regime remain mostly unchanged. However, the relative energy spacing of the states are somewhat affected in the interfacial confinement regime. This effect is maximum with full metallic type screening. It must be mentioned that a full metallic type screening is unrealistic in real MOSFETs as it forms a metal semiconductor junction allowing current leakage. The more realistic type of screening is likely to be of $\textrm{SiO}_2$ or of partial metallic nature. 

A plot of the valley splitting with screening in Fig 10 shows that valley splitting can vary by several meVs depending on the type of screening. This is expected as the electron image term varies rapidly near the interface boundary and modifies the confinement potential. 
 
 \begin{figure}[htbp]
 \centering
 \includegraphics[width=3in,height=2.4in]{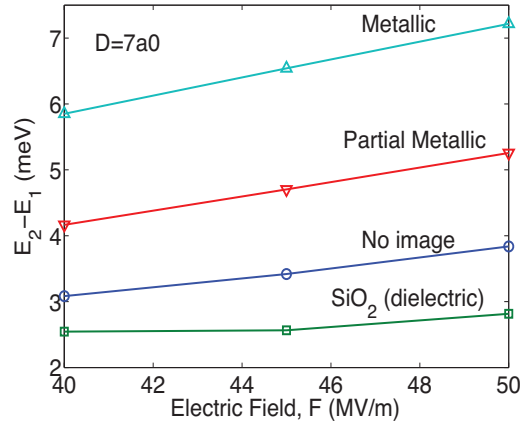}
 \caption{The splitting of the two lowest states at interfacial confinement as a function of field. This valley-orbit splitting can be sensitive to the type of screening.}
\label{fi:10}
\end{figure}
   
\section{IV. Conclusion}

We have computed the Stark shifted spectrum for an As and a P donor in Si at various depths from an interface. Utilizing the tight-binding approximation, we capture a more complete manifold of s and p type states skipped in earlier works. Understanding the details of these excited states, has proved to be critical to interpret experimental data {\cite{Rogge.NaturePhysics.2008}}. The results show adiabatic ionization of the donor electron to the field defined interfacial well as the higher excited states interfere with the 1s manifold. Anti-crossing between the ground state and the second excited state characterizes this ionization as the electron transits from a Coulomb confined regime to an interfacial confinement regime through an intermediate hybridization of the donor and interface well states. At weak field, surface effects are visible as the triplet and doublet degeneracy of the 1s manifold are lifted. At high fields, the states conform to the 2D symmetries of the interface well. Strong confinement by a hard wall and the field produces valley splitting of the lowest two states. Finally, we investigate the effect of various types of screening on the Stark shifted spectrum. We observe that the donor image term has a dominant effect in the Coulomb regime, while the electron image term is dominant in the interfacial confinement regime. 

The model and the method presented here helps to obtain a comprehensive quantitative description of the donor Stark shift problem that has been of recent interest in the context of quantum computing applications. This numerical approach helps in the large-scale qubit device modeling, and was used to interpret single donor transport measurements in FinFETs {\cite{Rogge.NaturePhysics.2008}}.   

\begin{acknowledgments}
This work was supported by the Australian Research Council, the
Australian Government, and the US National Security Agency (NSA) and the
Army Research Office (ARO) under contract number W911NF-08-1-0527. Part of the development of NEMO-3D was initially performed at JPL, Caltech under a contract with NASA. NCN/nanohub.org computational resources were used in this work. S.R. also acknowledges the support of Dutch Foundation for Fundamental Research on Matter (FOM) and the EU FP7 project AFSID. 
\end{acknowledgments}
Electronic address: rrahman@purdue.edu, lloydch@unimelb.edu.au


\begin{thebibliography}{100}
\bibitem{nanohub.note} nanoHUB.org computational resource of a
256-node 3.3GHz Pentium Irvindale PC cluster was used in this work. The tight-binding calculations were done under the hood of the Nano-Electronic Modeling Tool (NEMO-3D).
\end{thebibliography}
\end{document}